\documentstyle[twocolumn,prl,aps]{revtex}

\begin{document}

\twocolumn[\hsize\textwidth\columnwidth\hsize\csname
@twocolumnfalse\endcsname

\title{Differentiability Implies Continuity in Neuronal Dynamics}

\author{Joseph T. Francis$^{1,4}$ , Paul So$^{1,2}$, Bruce J. Gluckman$^{1,2}$, and \cite{corresponding}Steven J. Schiff$^{1,3,4}$}

\address{
$^1$Krasnow Institute for Advanced Studies, $^2$ Department of Physics and Astronomy, $^3$ Department of Psychology, George Mason University, George Mason University, Fairfax, VA, USA \\ $^4$ Program in Neuroscience, The George Washington University School of Medicine, Washington, DC.
}
\date{\today}
\maketitle

\begin{abstract}
Recent work has identified nonlinear deterministic structure in neuronal dynamics using periodic orbit theory. Troublesome in this work were the significant periods of time where no periodic orbits were extracted - ``dynamically dark" regions. Tests for periodic orbit structure typically require that the underlying dynamics are differentiable. Since continuity of a mathematical function is a necessary but insufficient condition for differentiability, regions of observed differentiability should be fully contained within regions of continuity. We here verify that this fundamental mathematical principle is reflected in observations from mammalian neuronal activity. First, we introduce a null Jacobian transformation to verify the observation of differentiable dynamics when periodic orbits are extracted. Second, we show that a less restrictive test for deterministic structure requiring only continuity demonstrates widespread nonlinear deterministic structure only partially appreciated with previous approaches.

PACS 5.45.Tp; 87.17.Nn; 87.18.Sn
\end{abstract}
]

Much progress has been made in recent years detecting deterministic dynamics within noisy experimental environments \cite{Kantz and Schreiber}. Traditionally, counting measures such as interspike interval histograms have been used to examine the distributions of neuronal firing times, and models of activity have assumed that fluctuations about the mean rates are stochastic \cite{Tuckwell}. Since both action potential generation and synaptic transmission are highly nonlinear processes, it is reasonable to assume that some of the fluctuations about neuronal mean firing rates are not generated from linear stochastic dynamics. Indeed, we have come to recognize that fluctuations in spike timing may appear deterministic, and that stochastic models may fail to account for neuronal timing \cite{Mainen and Sejnowski}. If neurons do indeed generate such detailed patterns of activity, our concept of how neuronal networks process information may require radical reinterpretation \cite{Hopfield}.

Gaining a better understanding of the behavior of a complicated dynamical system, which is differentiable near periodic orbits, has a celebrated history that reaches back over a century in celestial \cite{Hill}, and quantum mechanics \cite{Gutzwiller}. Chaotic systems possess an infinite number of invariant unstable periodic orbits \cite{Cvitanovic}. Although characterizing an infinite number of periodic orbits is impossible for real systems, there are three important features of the shorter orbits: 1) they are the ones most experimentally accessible, 2) the long term average behavior of such systems can be well approximated through them \cite{Artuso}, and 3) they can be used for predicting \cite{Pawelzik} and controlling such systems \cite{control,Schiff}.

It has recently been shown that periodic orbits can be identified within the neuronal activities from a broad range of hierarchical neuronal organization, from single cells through large ensembles \cite{So1998,Le Van Quyen}. A still unresolved issue for the extraction of periodic orbits from neuronal systems \cite{So1998,Pei and Moss}, are the ``dynamically dark" regions - significant time intervals where no periodic orbits seem to exist. We would anticipate that such highly structured dynamical features such as periodic orbits would not arise from a completely stochastic sea of activity, and then disappear suddenly. Indeed, since periodic orbits are hallmarks of deterministic activity, it is logical to presume that they arise from an underlying deterministic process, and become visible whenever either the noise within the system transiently decreases, or when the system approaches a particularly easy orbit to observe (e.g. an especially short one). We hypothesize that in the process of detecting periodic orbits, the apparently empty periods between resolved orbits \cite{So1998,Pei and Moss} contain structure that remains unobserved. We seek an explanation for this dynamical ``dark matter".

A common feature shared among many of the recently developed methods for the extraction of periodic orbits from experimental data is the implicit assumption that the underlying dynamics are differentiable \cite{methods,So19961997}. If the dynamics are differentiable it is a necessary condition that the underlying function is continuous \cite{Sprecher}. Periods of detected differentiability should therefore be fully contained within periods of continuity. Since continuity is a less restrictive criterion for determinism than differentiability (and should be easier to observe experimentally), we predict that regions of observed differentiability should arise from a more extensive background of continuity. Continuity should shed light on dynamically dark regions. 

Transverse hippocampal slices 350-400 $\mu$m thick were prepared from rat brain using techniques previously described \cite{Schiff}. Slices were perfused in an interface chamber at 3 ml/min with artificial cerebrospinal fluid (ACSF, composed of, in mM, 155 Na$^+$, 136 Cl$^-$, 3.5 K$^+$, 1.2 Ca$^{2+}$, 1.2 Mg$^{2+}$, 1.25 PO$_4^{2-}$, 24 HCO$_3^{-}$, 1.2 SO$_2^{-4}$, and 10 dextrose), at 34.5-35.1 $^o$C. Extracellular field potentials were recorded differentially using saline-filled micropipettes (2-4M$\Omega $). After a 90 min acclimation period the perfusate was changed to one containing one of four high [K$^+$] concentrations: 7.5mM, 8.5mM, 9.5mM, or 10.5mM, inducing spontaneous population burst-firing events within the network \cite{Rutecki}. A typical example of such stereotyped neuronal population burst-firing is shown in Fig. 1. The times of onset of these events were extracted, and the interevent time intervals between burst-firings used for analysis. All animal experiments were approved by the institutional animal care and use committee of George Mason University.

Delay coordinate embedding \cite{embedding} was used to reconstruct the underlying dynamics from these interburst intervals. To extract the unstable periodic orbits from reconstructed state space, a transformation based on observed local dynamics was applied \cite{So1998,So19961997} that concentrates the data density near the periodic orbits. In the limit of noise free dynamics the probability distribution function of the transformed data will be singular at the periodic points \cite{So19961997}. This reduces extraction of periodic orbits from experimental data to searching for peaks in the distribution of the transformed data. The concentration effect of this transformation, a ``dynamical lens", depends implicitly on the existence of differentiable dynamics in the underlying dynamical function. 

To illustrate why this is so, we use the transformation to extract a period one orbit from the one-dimensional discrete-time dynamical system $x_{n+1}=f(x_n)$, where $f(x_n)$ is the function that prescribes the evolution of the system state $x_n$ (in practice we use higher dimensional state space and dynamics, and extract higher period orbits \cite{So1998}). We assume that the dynamics described by the function $f(x)$ is unknown to us but that the local behavior of the dynamics, $f'(x_n)={{df(x_n)} \mathord{\left/ {\vphantom {{df(x_n)} {dx}}} \right. \kern-\nulldelimiterspace} {dx}}$, can be estimated from a local least squares fit to the experimental data. The transform $g(x_n,\kappa )$ of $x_n$ for period-1 orbits $x^*:f(x^*)=x^*$ is then defined as $g(x_n,\kappa )\equiv {{x_n-s(x_n,\kappa )\cdot x_{n-1}} \over {1-s(x_n,\kappa )}}$, where $s(x_n,\kappa )=f'(x_n)+\kappa \cdot (x_n-x_{n-1})$ is a function defined by the estimated local dynamics $f'(x_n)$ and by $\kappa $, an adjustable parameter of the transform. In practice, we use an ensemble average of densities of $g(x_n,\kappa )$, made with different randomly chosen values for $\kappa $. Significance of the observed peaks is then estimated by comparison with the transformed densities of Amplitude Adjusted Fourier Transform surrogates \cite{Theiler}, via the statistics of extremes \cite{Gumbel}. In particular, we estimate the probability that observed peaks in the transformed density are larger than the largest peaks observed in the surrogates. We employed between thirty and fifty surrogates for the analyses presented here, and only accepted peaks that lay entirely outside the distribution of maximal surrogate peaks. We estimate a resulting confidence limit of falsely identifying a periodic orbit of less than 3\%.

Note that peaks in the transformed density can be attributed to two main factors: first from a high density of points in the original data in recurrent positions - points that already appear periodic; and second from the concentrating effect of the transformation - points that map under the transformation closer to the periodic points. The latter effect requires successful estimation of the local derivatives of the function. A modification of this transformation that only detects recurrent positions is achieved by setting the terms corresponding to the estimates of the local derivatives to zero - a null Jacobian transformation \cite{Gluckman}. If, in the above one-dimensional analysis, we set $f'(x)=0$, the transformation reduces to $g(x_n,\kappa )=x_{n-1}+{1 \over {\left( {x_n-x_{n-1}} \right)^{-1}-\kappa }}$. We see that for recurrent points, i.e. $\left( {x_n-x_{n-1}} \right)\to 0$, the result approaches $g(x_n,\kappa )\to x_{n-1}$ , independent of $\kappa $. On the other hand, for $\left( {x_n-x_{n-1}} \right)$ large, $g(x_n,\kappa )\to x_{n-1}-{1 \over \kappa }$, which yields points broadly distributed as $\kappa $ is varied. By comparing the densities from these two transformations we distinguish whether peaks in the transformed density occur due just to recurrence in the original data, an aspect not well preserved by the surrogates, or are enhanced due to reasonable estimates of the local derivatives of the function. This null Jacobian transformation requires no surrogate data for validity, and is independent of the continuity test applied below. 

If the dynamics of these neurons could be accounted for by a mean rate with stochastic fluctuations, a linear stochastic model derived from the autocorrelation analysis would best fit our data \cite{Box and Jenkins}. The frequency of periodic orbit detections in each experimental group is indicated in Table 1. At higher levels of [K$^+$], all of the experiments demonstrated periodic orbit structure that could not be accounted for by linear properties.

In 100\% of cases where we detected periodic orbits, our estimated Jacobian matrices enhanced the density of points compared with null Jacobian transformations (Table 1). The typical amplification of the density peaks through the use of the estimated rather than null Jacobian matrices was approximately 5 (range 1.5 - 2100). Fig. 2 illustrates an example of the concentration effect of these estimated Jacobian matrices. These results help validate the presence of meaningful differentiable dynamics associated with our identified periodic orbits.

We next applied a test for functional continuity to these data \cite{Kaplan}. Continuity of a function requires that nearby points in the domain of definition of a function will be mapped close to each other in its range \cite{Sprecher}. In a complicated linear or nonlinear function, with or without noise, it may be difficult to see the close evolution of nearby trajectories since such close encounters may be rare or ``exceptional" \cite{Kaplan}. Nevertheless, when found these closely tracking points unmistakably reveal that a deterministic process is at hand. We employ a method which is sensitive enough to detect the (piecewise) continuity in a computer ``random" number generation algorithm in common use \cite{Kaplan}. If we have truly found differentiability in certain data windows, indicated by the existence of periodic orbits, these regions must be subsets of regions of continuity that are at least as large as the differentiable regions. Moreover, although regions of observed differentiability should frequently be contiguous with regions of continuity, the converse should not be valid; that is, regions of continuity might stand alone without contiguous regions of detected differentiability.

Following \cite{Kaplan} we searched for all pairs of points in reconstructed state space less than $\delta $ distance apart, $\left\| {X_n-X_m} \right\|<\delta ,\,\,n\ne m$, and then measured the distance $\varepsilon$ between the images of these pairs of points one interval into the future, $\left\| {X_{n+1}-X_{m+1}} \right\|=\varepsilon$. The average $\varepsilon$ for a given $\delta$, $\bar \varepsilon (\delta )$, is $\bar \varepsilon (\delta )={1 \over N}\sum\limits_{\scriptstyle {n,m}\hfill\atop
  \scriptstyle {n\ne m}\hfill}^{} {\left\| {X_{n+1}-X_{m+1}} \right\|}\,\,\,$, for all $n,m$ such that $\left\| {X_n-X_m} \right\|<\delta$, where N is the total number of pairs satisfying the inequality $\left\| {X_n-X_m} \right\|<\delta $, and $\left\| {\,\bullet \,} \right\|$ is the Euclidean norm. We then generated 100 surrogate data sets and determined how many surrogate sets have an $\bar \varepsilon $ less than or equal to the $\bar \varepsilon $ obtained for the experimental data at a given $\delta $. One-tailed parametric estimates of confidence limits are generated by the percentage of surrogate results with smaller $\bar \varepsilon $ than the experimental data.

Significant continuity was identified in all data windows with significant periodic orbits (Table 1). The surrogate data test for significance makes it unlikely that such evidence for functional continuity could be accounted for by the linear stochastic properties of the intervals. As predicted above, we observed that regions with differentiability, containing periodic orbits, were subsets of and therefore contiguous with regions where only continuity was found. Furthermore, regions with highly significant continuity were found widespread, not necessarily associated with differentiable areas. As expected, regions with differentiability without continuity were not encountered. Whereas significant periodic orbits were detected in 15-24\% of the data time windows, the length of time occupied by significant continuity in these experiments was more than twice this amount.  A typical example of the intermingling of regions with continuity and differentiability from these experiments is illustrated in Fig. 3.

These data provide compelling evidence that the fluctuations about the mean burst rate of these neuronal networks are frequently deterministic. Periodic orbits reveal only a fraction of the amount of nonlinear deterministic behavior detectable in these systems. The degree of nonlinear deterministic activity increased as [K$^+$] increased. We made these observations despite having used neuronal networks cut off from their natural inputs and stimulation patterns, perhaps the worst condition for detecting structure in neuronal dynamics \cite{de Ruyter van Steveninck}.

There are many features in biological systems that limit the resolvability of differentiability and hence periodic orbits. Biological systems are nonstationary, and this is reflected in our data in the disparity between the percentage of experiments (50-100\%) versus percentage of data windows (15-24\%), within which periodic orbits were identified (Table 1). Nevertheless, our use of relatively small data windows (128 interburst intervals, ~2-3 minutes of data) reduces the effect of nonstationarity on our analysis. On the other hand, our ability to resolve differentiable maps from experimental systems will be constrained by such finite sampling of the underlying dynamical function. We therefore remain cautious in equating our ability to resolve periodic orbits with existence of differentiability in the underlying dynamics. Nevertheless, our null Jacobian results are fully consistent with the observation of differentiability from these experiments.

Our dynamical ``dark matter" is therefore revealed to contain functions drawn from the set of continuous (or piecewise continuous) mathematical functions. In addition, our findings lend further validity to previous findings of periodic orbits within such dynamics. From a practical point of view in handling experimental data, continuity is a less restrictive feature in the functional representation of a dynamical system than is differentiability, and is easier to detect. If the goal is to demonstrate determinism in experimental data, continuity appears to be a better test than periodic orbit extraction. 

 Acknowledgements: We thank T. Sauer and T. Netoff for helpful discussions. Supported by NIH 7K02MH01493 and 2R01MH50006 and NSF IBN 97-27739.

\begin{tabular}{l|l|l|l|l|l|l}
\lbrack K$^{+}$] & N & \%N  & W & \%W  & \% UPOs & \%W with \\ 
&  & with  &  & with &  with & UPOs and \\ 
&  & UPOs &  & UPOs & Enhancement & Continuity \\ \hline
7.5 & 8 & 50 & 33 & 15 & 100 & 100 \\ 
8.5 & 9 & 77 & 44 & 16 & 100 & 100 \\ 
9.5 & 11 & 82 & 71 & 23 & 100 & 100 \\ 
10.5 & 11 & 100 & 68 & 24 & 100 & 100
\end{tabular}

\begin{table}
\caption{Summary of experimental results.\lbrack K$^{+}$] = potassium concentration in perfusate; N = number of experiments; W = number of data windows; UPOs = unstable periodic orbits; Enhancement indicates enhanced density about candidate UPOs. To enhance sensitivity, and compensate for nonstationarity, periodic orbits were identified using overlapping 128 point data windows. Statistics in Table 1 are reported for non-overlapping windows in order to ensure that windows and periodic orbits were not over counted.}
 \label{table}
\end{table}

\begin{figure}
\caption{Typical extracellular recording of voltage deflections corresponding to network burst firing in a population of neurons. Calibration bars: voltage (mV) and time (sec). Intervals $I_i$ determined from onset times as shown.}
\label{figure1}
\end{figure}

\begin{figure}
\caption{Example of transformed data density histogram used to isolate periodic orbits. Thick solid line is experimental result using full transform (recurrence plus differentiability), thin solid line is experimental density from transform using null Jacobian (recurrence), and dotted line is mean of the surrogate data densities using full transform. Data from experiment at 10.5 mM [K$^+$]. Highest peak indicates periodic orbit. The increase in amplification of this peak due to the estimated Jacobian matrix was approximately four-fold, typical of results. This amplified peak lay outside of the distribution of surrogate peaks.}
\label{figure2}
\end{figure}

\begin{figure}
\caption{Simultaneous comparison of observation of periodic orbits (indicated by triangles) with significant continuity (gray scale). Abscissa indicates window number for experiment, equivalent to time. Ordinate indicates values of $\varepsilon$ in arbitrary units. Gray scale represents probability that functional continuity is significant, with dark gray revealing less than 5\% chance of finding continuity in the surrogates. Since these statistics are on a pixel by pixel basis, confluently gray regions represent continuity with much greater than 95\% confidence. Data windows represent 128 interspike intervals, with neighboring windows overlapped 126 intervals. The two periodic orbits indicated were period-1 orbits, centered about 960 and 985 msec respectively. Note that each periodic orbit is contained in time within a larger period of continuity, and that substantial regions of continuity are present within which no differentiability is identified.}
\label{figure3}
\end{figure}

\end{document}